\documentstyle[12pt]{article}
\def\beq{\begin{equation}}
\def\eeq{\end{equation}}
\def\beqa{\begin{eqnarray}}
\def\eeqa{\end{eqnarray}}
\begin{document}
\begin{flushright}
UH--511--919--98 \\
November 1998
\end{flushright}
\begin{center}
\Large
{\bf Unexpected symmetries in classical moduli spaces}
\end{center}
\normalsize
\bigskip
\begin{center}
{\large  Tonnis A. ter Veldhuis
}
\\
{\sl Department of Physics \& Astronomy\\
University of Hawaii\\
Honolulu, HI 96822, USA}\\
\vspace{2cm}

\end{center}
\centerline{ABSTRACT}
\vspace{1cm}
\noindent
 
We analyze the structure
of the moduli space of a supersymmetric $SU(5)$ chiral gauge theory 
with $2$ matter fields in the ${\bf 10}$ representation, and 
$2$ fields in the ${\bf \bar{5}}$ representation. 
Inspection of the exact K\"{a}hler potential
of the classical moduli space shows that the symmetry group of the  moduli
space is larger than the global symmetry group of the
underlying gauge theory. As a consequence, the gauge theory
has classical inequivalent vacua which yield identical low energy 
theories.
\\
\vfill\eject

\section{Introduction}

Depending on the matter content, supersymmetric gauge
theories can have  large vacuum degeneracies \cite{BDSF}.
In the absence of a superpotential, classical vacua 
are associated with vacuum expectation values for
which the D--terms of the scalar potential vanish.
In Wess--Zumino gauge, the
D--flat directions contain those points in the vector space
of scalar components of the chiral superfields that satisfy the condition
\beqa
D^a & = & \sum_i \phi_i^{\dagger} T^a \phi^i = 0, \label{Dflat}
\eeqa 
where the sum is over all matter multiplets, $\phi^i$ is
the scalar component of the superfield $\Phi^i$, and $T^a$ are the 
generators of the gauge group in the appropriate representation. 
In case all matter transforms under 
(anti)--fundamental representations of the gauge group, it is relatively
simple to construct
solutions to Eq. (\ref{Dflat}), 
but for theories with matter in tensor representations,
the solutions may be rather complex. No standard methods
to find the most general solution are available in the latter case.
Ref. \cite{Shi} gives an overview of the efforts to parametrize
flat directions in various models.

The D-flatness condition
Eq. (\ref{Dflat}) is covariant under the gauge group $G$
and invariant under the global symmetry group $H_G$ of the gauge
theory. The manifold of flat directions is therefore covered
with $G \otimes H_G$ orbits.  Points in the manifold 
that lie on the same $G \otimes H_G$ orbit are physically equivalent.  
The analysis of the flat directions is therefore simplified considerably when
the redundancy due to gauge and global symmetry transformations is removed.

To this end, the $G$ orbits in the flat direction manifold can labeled
by a finite set of basic holomorphic gauge invariant polynomials
$X_n(\phi^i)$ \cite{ADS,ADS1,LT}. Any holomorphic gauge invariant polynomial 
in the fields $\phi^i$ can be written 
in terms of products and sums of the basic invariants $X_n$,
by virtue of the decomposition rules for the products of
representations of the fields $\phi^i$.
For some theories the invariants $X_n$ are algebraically independent;
for others, relations exist among them.

The invariants $X_n$ form the coordinates of the moduli space. (the
flat direction manifold modulo gauge transformations)
The $H_G$ orbits in the moduli space can be labeled by the 
finite set $\{I_x\}$
-- the basic, $H_G$ invariant, Hermitian polynomials in terms of $X_n$
and $X_n^{\dagger}$.
The moduli space is, in fact, a K\"{a}hler manifold.
Its K\"{a}hler potential, induced by the K\"{a}hler potential of the
gauge theory, is defined by $K_M(I_x(X_n^{\dagger},X_n))=
\phi_i^{\dagger} \phi^i$ for every point in the flat direction
manifold.

When the $X_n$ are promoted to chiral superfields,
the moduli space becomes equivalent
to a supersymmetric chiral sigma model \cite{ADS1,PR}. This
sigma model describes the low
energy limit of the underlying gauge theory 
if the gauge symmetry is completely broken -- 
the effective, classical theory describing the low energy limit of
the gauge theory built on the classical
vacuum  with expectation values $<\phi^i>$
is equivalent to the sigma model
when it is expanded around the expectation values $<X_n>=X_n(<\phi^i>)$.
This effective theory, which describes
the interactions of the massless 
degrees of freedom, can also be obtained directly by integrating
out the massive vector multiplets\footnote{
At the classical level, this means
considering tree--diagrams with only massless degrees
of freedom at the external lines, and contracting internal
propagators of massive degrees of freedom  in the
limit $p^2/M^2 \rightarrow 0$.} in the gauge theory.

Non-perturbative effects can change
the classical picture of the moduli space dramatically \cite{Shi,IS,Pes}. 
In some
cases, a dynamically generated superpotential lifts the
moduli space; in other cases, classical constraints
among the moduli fields are modified; and in still other cases
the structure of the moduli space remains unchanged.
Holomorphy and symmetries severely constrain the form
of dynamically generated superpotentials. Unfortunately, modifications
to the K\"{a}hler potential are less well understood.

In some situations, however, corrections to the classical K\"{a}hler potential
are  small.
This is for example the case in models
with calculable dynamical supersymmetry breaking, where
the vacuum expectation values of the scalar fields
are much larger than the dynamical scale of the
gauge theory.  

By construction, the K\"{a}hler potential of the classical moduli space is
invariant under global symmetry group $H_G$ of the underlying 
gauge theory.
By a detailed analysis of the moduli space of
a chiral $SU(5)$ gauge theory with two
antisymmetric tensors and two anti-fundamentals,
we will show that
the symmetry group $H_M$ of the moduli space
can be larger than $H_G$.
When a superpotential is added and non--perturbative
effects are taken into account, this $SU(5)$ theory is one of the
classic models with calculable dynamical supersymmetry breaking
\cite{Shi,ADS2,MV,VZS,TtV1,TtV2}.

The fact that $H_M$ is larger than $H_G$ has some
interesting consequences. 
Classical vacua which are not related by gauge and global
symmetry transformations still give rise to the same
effective theory in the low energy limit. Moreover, the
unbroken symmetry group in the effective theory extends
the unbroken symmetry group of the full gauge theory.
We have verified that the extended
symmetry is not a consequence of a custodial symmetry. Moreover,
as we calculated the {\it exact} classical K\"{a}hler potential,
it is not a consequence of a truncation either.

In Section 3, we present a detailed analysis of the
moduli space of the $SU(5)$ theory. However, we first discuss a
simple, well-known, vector-like theory with $SU(3)$ gauge
symmetry \cite{ADS1} in Section 2; this  serves
to illustrate our methods,  and to emphasize the main point 
of this Letter by contrast -- as the symmetry group
of the moduli space of the $SU(3)$ model coincides with the global symmetry
group of the underlying gauge theory.

\section{Supersymmetric QCD with two flavors}

We consider supersymmetric
QCD with three colors and two flavors \cite{ADS1}.
The quark chiral superfields, which are denoted by $Q_a^i$, 
and $\bar{Q}_i^{\alpha}$, transform
as ${\bf 3}$ and ${\bf \bar{3}}$ under $SU(3)$. 
Here $i=1,2,3$ is the color index, and $a=1,2$ and
$\alpha=1,2$ are flavor indices. The global symmetry
group\footnote{
According to our conventions for R-symmetry,
the charge of the
scalar component of a chiral superfield is R, whereas the charge of the 
fermionic component is $R-1$. The gaugino has charge $1$.} 
$H_G$ of the theory --
the relevant symmetry group at the classical level --
is $SU(2)_Q \otimes SU(2)_{\bar{Q}} 
\otimes U(1)_Q \otimes U(1)_{\bar{Q}} \otimes U(1)_R$.
Under $H_G$ the quark superfields transform as 
$Q_a^i \sim (2,1,1,0,0)$ and $\bar{Q}_i^{\alpha} \sim (1,2,0,1,0)$.
The scalar components of the chiral superfields do not transform
under $U(1)_R$, and therefore this factor can not be spontaneously
broken by expectation values of the scalar fields. 

The non-anomalous subgroup $H_{NA}$ of $H_G$ --
the relevant symmetry group at the quantum level -- is
$SU(2)_Q \otimes SU(2)_{\bar{Q}} \otimes U(1)_B \otimes U(1)_{R'}$.
Under $H_{NA}$ the quark superfields transform as 
$Q_a^i \sim (2,1,1,-1/2,)$ and $\bar{Q}_i^{\alpha} \sim (1,2,-1,-1/2)$.

The flat directions of the theory are solutions to the equation
\beqa
Q_i^{\dagger a} Q_a^j - \bar{Q}_{\alpha}^{\dagger j} \bar{Q}_i^{\alpha}
& = & c \delta_i^j. \label{QCDflat}
\eeqa
Here $c$ is, a priori, an arbitrary real constant. However,
it turns out there are only solutions for $c=0$. 
Any solution to Eq. (\ref{QCDflat})
can be obtained from the solution  $Q_1^1=\bar{Q}_1^1=a$, 
$Q_2^2=\bar{Q}_2^2=b$, with
all other components vanishing, by applying appropriate gauge and global 
symmetry transformations.
For generic values of the real parameters $a$ and $b$, the gauge group 
is completely 
broken. Eight of the twelve chiral superfields are  eaten
to give mass to the vector multiplets. As a consequence,
the number of moduli fields is four and the moduli space is eight--dimensional.
The unbroken global symmetry group is $U(1) \otimes U(1) \otimes U(1)_R$.
The moduli space is therefore spanned by the two parameters $a$ and
$b$, and six of the nine parameters of $H_G$ transformations.

The basic holomorphic gauge invariants for this theory are
$M_a^{\alpha} = \bar{Q}_i^{\alpha} Q_a^i$,
transforming as $(2,2,1,1,0)$ under $H_G$. These four meson fields form
the coordinates of the moduli space, and their vacuum
expectation values can be written
in the form $M_1^1=m_1$, $M_2^2=m_2$, and $M_1^2=M_2^1=0$, by
$H_G$ transformations. 

The basic Hermitian
structures --  invariant under the global symmetry
transformations and constructed out of the 
meson fields -- are 
$I_1=M_{\alpha}^{\dagger a} M_a^{\alpha}$
and
$I_2=M_{\alpha}^{\dagger a} M_{\beta}^{\dagger b }M_a^{\beta} M_b^{\alpha}$,
with the range of $I_2$ limited by the inequality
$1/2 I_1^2 \leq I_2 \leq I_1^2$.
The {\it exact} induced K\"{a}hler potential of the 
classical moduli space is defined
as $K_M(I_1(M^{\dagger},M),I_2(M^{\dagger},M)) \equiv Q_i^{\dagger a} Q_a^i +
\bar{Q}_{\alpha}^{\dagger i} \bar{Q}_i^{\alpha}$, and
a simple calculation gives
\beqa
K_M & = & 2 \sqrt{\frac{1}{2} I_1 + \frac{1}{2} \sqrt{2I_2 -I_1^2}} + 
  2 \sqrt{\frac{1}{2} I_1 - \frac{1}{2} \sqrt{2I_2 -I_1^2}}. \label{Kmod}
\eeqa
This K\"{a}hler potential is invariant under the global symmetry group
$H_G$ of the underlying gauge theory by construction. 
As is conventional, $K_M$ is not invariant under any other symmetries, so that
the symmetry group $H_M$ of the moduli space  is equal to $H_G$. 
As will become clear in the next
section, however, even though $H_M$ always contains $H_G$, it can in fact
be larger. 

The K\"{a}hler potential of the moduli space is derived in terms
of the scalar components of the superfields. However, when the
the moduli fields $M_a^{\alpha}$ are promoted to superfields, a
supersymmetric sigma model ensues. The low energy limit of the
classical gauge theory constructed on the vacuum with expectation
values $<Q_a^i>$ and $<\bar{Q}_i^{\alpha}>$ is equivalent to the sigma model
with vacuum expectation values $<M_a^{\alpha}>=<\bar{Q}_i^{\alpha}><Q_a^i>$.

The $H_G$ orbits that cover the moduli space can be labeled by 
$\{a,b\}$, or $\{m_1,m_2\}$, or $\{I_1,I_2\}$. Points in the moduli
space that lie on the same orbit yield physically equivalent classical
vacua. The orbits, in turn, can be grouped into strata. Different orbits 
that belong to the same stratum yield vacua that are physically inequivalent,
but qualitatively similar. Such vacua yield
the same symmetry breaking pattern and the same degeneracies
in the mass spectrum, but the masses are quantitatively different.
The strata can be categorized as follows:
\begin{itemize}
\item[{\it i})]{ For generic $H_G$ orbits, labeled by generic values of
$\{I_1,I_2\}$, the little group is $U(1) \otimes U(1)
\otimes U(1)_R$. The gauge symmetry is completely broken.} 
\item[{\it ii})]{ 
For orbits with $I_2 = 1/2 I_1^2$, ($b=\pm a$; $m_1=\pm m_2$) the little group 
is $SU(2) \otimes U(1) \otimes U(1)_R$. 
The gauge symmetry is completely broken.}
\item[{\it iii})]{For orbits with  $I_2=I_1^2$, ($b=0$; $m_2=0$) the little group is 
$U(1)\otimes U(1) \otimes U(1) \otimes U(1)_R$.
In the sigma model, the metric derived from the K\"{a}hler potential
$K_M$  is singular.
Moreover, in the gauge theory the gauge group is  broken to $SU(2)$ 
and therefore, the low energy theory should include the massless 
gauge multiplets.}
\item[{\it iv})]{When $I_1=0$, none of the gauge and global symmetries 
are broken.}
\end{itemize}

The classical picture of the moduli space is altered dramatically
by non-perturbative effects. 
The non-anomalous global symmetry group $H_{NA}$ of the gauge theory
allows a unique, non-perturbative superpotential \cite{ADS1} of the form
\beqa
W_{np} = \frac{\Lambda^7}{
\bar{Q}_i^{\alpha} Q_a^i \bar{Q}_j^{\beta} Q_b^j 
\epsilon_{\alpha \beta} \epsilon^{ab}}.
\eeqa
Explicit instanton calculations in 
the semi-classical approximation \cite{Cor,FP} show that
such an effective superpotential is indeed generated
and that $\Lambda$ is the dynamical scale of the gauge theory.
The F-term contributions to the scalar potential completely lift the D--flat
directions.  The scalar potential does not have a minimum, tends
to zero only at infinity, and renders the theory unstable. 
However, the scalar potential is stabilized if a mass term of the form
\beqa
W_m & = & m_{\alpha}^a \bar{Q}_i^{\alpha} Q_a^i
\eeqa
is added to the superpotential. If the scale of the masses $m_{\alpha}^a$ is
much smaller than the dynamical scale $\Lambda$, then the vacuum expectation
values of the scalar fields are much larger than $\Lambda$ and the
theory is weakly coupled. It is in this limit that the
classical K\"{a}hler potential is relevant. 
The theory below the dynamical scale can
be described in terms of the moduli fields $M_a^{\alpha}$, with
K\"{a}hler potential $K_M$ and superpotential
\beqa
W & = & \frac{\Lambda^7}{M_a^{\alpha} M_b^{\beta} \epsilon_{\alpha \beta}
\epsilon^{ab}} + m_{\alpha}^a M_a^{\alpha}.
\eeqa
The vacuum energy vanishes, and supersymmetry is not broken in this theory.

\section{Chiral $SU(5)$ theory}

The chiral
supersymmetric $SU(5)$ gauge theory we  discuss in this section
contains two matter fields transforming under the ${\bf 10}$ 
representation of $SU(5)$, and two fields transforming under
the ${\bf \bar{5}}$ representation. These matter fields are
denoted by the two index anti-symmetric tensors $T_a^{ij}$,
and $\bar{F}_i^{\alpha}$, where $i,j=1,..5$ are
gauge indices, and  $a=1,2$ and $\alpha=1,2$ are flavor
indices. With this matter content,
the theory is anomaly free and asymptotically free.

The global symmetry group $H_G$ of the theory is 
$SU(2)_T \otimes SU(2)_{\bar{F}} \otimes U(1)_{T} 
\otimes U(1)_{\bar{F}} \otimes U(1)_{R}$.
Under $H_G$, the  matter 
fields transform as $T_a \sim (1,2,1,0,0)$ and 
$\bar{F}^{\alpha} \sim  (2,1,0,1,0)$. The scalar
components of the chiral superfields do not
transform under $U(1)_R$. 
Their vacuum expectation values therefore
do not break this symmetry. 
Under the non-anomalous
subgroup of $H_G$, 
$SU(2)_T \otimes SU(2)_{\bar{F}} \otimes U(1)_{A} \otimes U(1)_{R'}$,
the matter fields transform as 
$T_a \sim (1,2,1,1)$ and 
$\bar{F}^{\alpha} \sim  (2,1,-3,-4)$.

The D-flat directions of the theory are solutions to the
equation
\beqa
T_{ij}^{a\dagger} T_{a}^{ik} - 
\bar{F}_{\alpha}^{k\dagger} \bar{F}_j^{\alpha} = c
{\delta}_j^k, \label{flateq}
\eeqa
where $c$ is an arbitrary real constant. In Refs. \cite{Shi,ADS2,VZS,TtV2},
some incomplete families of solutions to Eq. (\ref{flateq}) were presented.
Here, we give the most general solution  which  of course  includes
the previously--found families. Any solution to
Eq. (\ref{flateq}) can be obtained from a four--parameter solution
through gauge and global symmetry transformations. This four--parameter 
solution takes the form,
$T_2^{12}  =  a$,
$T_2^{34}  =  b$,
$\bar{F}_1^1  =  c$, 
$\bar{F}_5^1  =  d$, and
\beqa
T_1^{13} & = & \frac{c}{b}  
        \sqrt{a^2-c^2} \sqrt{\frac{b^2}{a^2-c^2}+\frac{d^2}{a^2}},
\nonumber \\
T_1^{45} & = & \frac{a}{b} 
        \sqrt{a^2-c^2} \sqrt{ \frac{b^2}{a^2-c^2}+\frac{d^2}{a^2}},
\nonumber \\
T_2^{23} & = & \frac{c}{\sqrt{a^2-c^2}} \sqrt{b^2-(a^2-c^2)},   
\nonumber \\
T_2^{25} & = & -\frac{c d}{a}, 
\nonumber \\
T_2^{45} & = & \frac{d}{b a} \sqrt{a^2-c^2} \sqrt{b^2-(a^2-c^2)}, 
\nonumber \\      
\bar{F}_3^1 & = & -\frac{a}{\sqrt{a^2-c^2}} \sqrt{b^2-(a^2-c^2)}, 
\nonumber \\
\bar{F}_2^2  & = & \frac{c}{b} \sqrt{b^2-(a^2-c^2)} 
        \sqrt{ \frac{b^2}{a^2-c^2}+\frac{d^2}{a^2}},  
\nonumber \\
\bar{F}_4^2 & = & - \sqrt{a^2-c^2} 
            \sqrt{ \frac{b^2}{a^2-c^2}+\frac{d^2}{a^2}}. \label{gensol}
\eeqa
All other components vanish, and $a$, $b$, $c$ and $d$ are real parameters.
For generic values of $\{a,b,c,d\}$, the gauge symmetry is completely
broken. Therefore, twenty-four of the thirty chiral superfields
are eaten to give masses to the vector multiplets, leaving
six moduli fields to function as coordinates for the 
twelve--dimensional moduli space.
The global symmetry group $H_G$
is broken to $U(1)_R$. In terms of the fundamental fields, the moduli 
space is spanned by the four
parameters $\{a,b,c,d\}$ of the solution given in Eq. (\ref{gensol}),
and eight of the nine parameters of $H_G$ transformations.
The basic holomorphic gauge invariants for this theory are
given by
\beqa
X_a & = & \epsilon_{\alpha \beta} \bar{F}_i^{\alpha} 
\bar{F}_j^{\beta} T_a^{ij},
\nonumber \\
J_a^{\alpha} & = & \epsilon_{ijklm} \bar{F}_n^{\alpha}
 T_a^{ij}  T_b^{kl} T_c^{mn} \epsilon^{bc}. \label{mf}
\eeqa
Under $H_G$, these holomorphic gauge invariants 
transform as $X_a \sim (1,2,1,2,0)$ and $J_a^{\alpha} \sim (2,2,3,1,0)$. 
By suitable $H_G$ transformations, the vacuum expectation values of
the basic holomorphic gauge invariants can be written as
$X_1=x_1$, $X_2=x_2$, $J_1^1=j_1$, $J_2^2=j_2$ and $J_1^2=J_2^1=0$, with
$x_1$, $x_2$, $j_1$ and $j_2$ real parameters.
In fact, the expectation values of the holomorphic gauge invariants 
for the four--parameter solution, given in Eq. (\ref{gensol}), already have this 
form:
\beqa
X_1 & = & 2 \frac{a d}{b} (a^2 -c^2)
            \left( \frac{b^2}{a^2-c^2}+\frac{d^2}{a^2} \right),
\nonumber \\
X_2 & = & 2 \frac{a^3}{b} \sqrt{b^2-(a^2-c^2)}  
            \left( \frac{b^2}{a^2-c^2}+\frac{d^2}{a^2} \right)^{3/2}, 
\nonumber \\
J_1^1 & = & 12 \frac{a^2 c^2}{b^2} (a^2-c^2)
               \left( \frac{b^2}{a^2-c^2}+\frac{d^2}{a^2} \right)^2, 
\nonumber \\
J_1^2 & = & 0, 
\nonumber \\
J_2^1 & = & 0, 
\nonumber \\
J_2^2 & = & -12 a^2 (a^2-c^2) 
                \left( \frac{b^2}{a^2-c^2}+\frac{d^2}{a^2} \right).                
\eeqa
The holomorphic invariants $X_a$ and $J_a^{\alpha}$ provide the coordinates 
for the moduli space.  A completely $H_G$ invariant description of the moduli 
space can be given in terms of the four Hermitian invariants
\beqa
I_1 & = & {{X}^a}^{\dagger} X_a,  \nonumber \\
I_2 & = & {J_{\alpha}^a}^{\dagger} J_a^{\alpha},  \nonumber\\
I_3 & = & {X^a}^{\dagger} {J_{\beta}^b}^{\dagger} X_b J_a^{\beta}, \nonumber \\
I_4 & = & {J_{\alpha}^a}^{\dagger} {J_{\beta}^b}^{\dagger} J_a^{\beta} 
J_b^{\alpha}, \label{globinv}
\eeqa
where the range of $I_4$ is limited to $1/2 I_2^2 \leq I_4 \leq I_2^2$,
and the range of $I_3$ is limited by 
$(2 I_3 -I_1 I_2 )^2 \leq (2 I_4 -I_2^2) I_1^2$.
The moduli space is thus covered by $H_G$ orbits,
labeled by  $\{a,b,c,d\}$, or $\{x_1,x_2,j_1,j_2\}$,
or $\{I_1,I_2,I_3,I_4\}$.
In our previous work \cite {TtV2}, the exact K\"{a}hler 
potential\footnote{
The K\"{a}hler potential of the moduli space 
$K_{M}(I_1,I_2,I_3,I_4)= 1/2 T_{ij}^{a\dagger} T_a^{ij} + 
\bar{F}_{\alpha}^{i \dagger} \bar{F}_i^{\alpha}$
for all values of the parameters $\{a,b,c,d\}$ of the four--parameter solution
to the D-flatness equation.}
of the classical moduli space was
derived. Invariance under $H_G$ dictates that the K\"{a}hler
potential has the functional form
\beqa
K_{M}(X^{\dagger},X,J^{\dagger},J) & = & K_{M}(I_1,I_2,I_3,I_4).
\eeqa
Defining
\beqa
A & = & 125 I_1 \nonumber \\
B & = & \frac{25}{9} 
\left( \sqrt{\frac{1}{2} I_2 + \frac{1}{2} \sqrt{2 I_4 - I_2^2}} +
       \sqrt{\frac{1}{2} I_2 - \frac{1}{2} \sqrt{2 I_4 - I_2^2}} \right),
\eeqa
and
\beqa
p & = & 2 \sqrt{B} 
\cos ( \frac{1}{3} \arccos \frac{A}{B^{\frac{3}{2}}} ),
\eeqa
the K\"{a}hler potential of the moduli space is given by
\beqa
K_{M}  & = & \frac{3}{10} \left( p + \frac{B}{p} \right). \label{kp}
\eeqa
The metric derived from this K\"{a}hler potential 
is singular if $I_4 = I_2^2$. Curiously, $K_M$
does not depend on $I_3$. 
As a consequence -- and this illustrates the central point of this Letter
-- the symmetry group $H_M$ of the 
moduli space,
$SU(2)_X \otimes SU(2)_1 \otimes SU(2)_2 \otimes U(1)_X \otimes U(1)_J
\otimes U(1)_R$,
is larger than the global symmetry group $H_G$ of the underlying 
gauge theory.  
The moduli fields transform under $H_M$ as $X_a \sim (2,1,1,1,0,0)$
and $J_a^{\alpha} \sim (1,2,2,0,1,0)$. The $U(1)_R$ factor in
$H_M$ is the same factor that appears in $H_G$. Only fermions
transform under this symmetry, and it does not play any role
in the discussion below. We will therefore suppress this factor
from here on.

Generic $H_M$ orbits in the moduli space, labeled
by $\{I_1,I_2,I_4\}$, contain one-parameter families
of $H_G$ orbits, labeled by $I_3$. 
In particular, the points
\beqa
X_1 & = & x \cos \phi, \nonumber \\
X_2 & = & x \sin \phi, \nonumber \\
J_1^1 & = & j_1 \nonumber, \\
J_1^2 & = & 0 \nonumber, \\
J_2^1 & = & 0 \nonumber, \\
J_2^2 & = & j_2, 
\eeqa
for fixed values of $\{x,j_1,j_2\}$, and varying $\phi$, are 
equivalent in the moduli space, as $\phi$ corresponds to the
parameter of an $SU(2)_X$ rotation. $H_M$ orbits can therefore also
be labeled by $\{x,j_1,j_2\}$, and the $H_G$ orbits contained
in an $H_M$ orbit can be labeled by $\phi$.

When the moduli fields are promoted to superfields, a supersymmetric
sigma model results.
The low energy limit of the gauge theory built on the classical
vacuum with expectation values $<T_a^{ij}>$ and $<\bar{F}_i^{\alpha}>$
is equivalent to the sigma model with vacuum expectation
values $<X_a>  =  \epsilon_{\alpha \beta} <\bar{F}_i^{\alpha}>
<\bar{F}_j^{\beta}> <T_a^{ij}>$ and 
$<J_a^{\alpha}>  =  \epsilon_{ijklm} <\bar{F}_n^{\alpha}>
<T_a^{ij}>  <T_b^{kl}> <T_c^{mn}> \epsilon^{bc}$.

The extended symmetry of the moduli space has two important
consequences. First, vacua of the gauge theory corresponding to fixed values
of $\{x,j_1,j_2\}$, but varying $\phi$, are physically inequivalent. 
In particular,
the masses\footnote{
The mass spectrum of the vector multiplets,
which we studied numerically, displays some unusual features. For
generic values of $\{x,j_1,j_2,\phi\}$ both the gauge and
global symmetries of the gauge theory are completely broken. However,
the spectrum contains four
degenerate pairs of masses, and one degenerate quintuplet. 
Moreover, even though the spectrum changes with $\phi$ for fixed
values of $\{x,j_1,j_2\}$, the sum of the squares of the masses
and the mass of the degenerate quintuplet remain independent of $\phi$.}
of the vector multiplets
are a function
of $\phi$. In fact, while for generic values
of $\phi$ gauge and global symmetries are completely
broken, for the special values of $\phi=0$ and $\phi=\pi/2$ there is a remaining
global $U(1)$ symmetry. However, all
vacua of the sigma model with fixed values of $\{x,j_1,j_2\}$ and 
arbitrary value of $\phi$,
either generic or special, are
equivalent. 
Therefore, the low energy limit of the gauge theory, which is 
obtained by integrating out the massive vector
multiplets in the limit $p^2/M^2 \rightarrow 0$, is identical
for each value of $\phi$. {\it Physically inequivalent vacua
of the gauge theory, with distinct mass spectra and possibly
even distinct global symmetry breaking patterns, yield the same
low energy theory.} Second, for generic vacua, the global
symmetry group of the gauge theory is broken to $U(1)_R$. 
However, the symmetry group $H_M$ of the moduli space is
broken to $U(1) \otimes U(1) \otimes U(1)_R$. Therefore, 
the low energy limit of the gauge theory has a larger
symmetry group than expected from the global
symmetry breaking pattern of the full gauge theory.

The three--parameter solution to the D-flatness condition
Eq. (\ref{flateq}), obtained by
imposing the condition $b^2=a^2-c^2$ on the four--parameter solution 
given in Eq. (\ref{gensol}),
corresponds to arbitrary $\{I_1,I_2,I_4\}$ and $I_3=0$, or, alternatively,
arbitrary $\{x,j_1,j_2\}$ and $\phi=0$.
This three--parameter solution, therefore, contains
a representative point on all $H_M$ orbits in the moduli space.
However, it does not contain a representative point on 
all $H_G$ orbits.
Therefore, the corresponding classical vacua 
yield all physically inequivalent low energy theories,
yet not all physically inequivalent classical gauge theories.

We will describe the moduli space in terms of strata of $H_M$ and
$H_G$ orbits in turn. The first approach lends itself for the study of all
inequivalent low energy theories, while the latter is more
suitable for the study of all inequivalent classical gauge theories.

As explained before, $H_M$ orbits are labeled by either
$\{I_1,I_2,I_4\}$ or $\{x,j_1,j_2\}$.
For generic orbits, labeled by generic 
values of $\{x,j_1,j_2\}$, $H_M$
is broken to $U(1)\otimes U(1)$. One of the $U(1)$ factors is a subgroup
of $SU(2)_X \otimes U(1)_X$; the other, a subgroup of 
$SU(2)_1 \otimes SU(2)_2 \otimes U(1)_J$. The number of 
broken symmetry generators, nine, is larger than the number of moduli
fields, six, and therefore some of the corresponding Goldstone bosons
are non-doubled.
Apart from the generic stratum, there are strata for which the  
little group is larger.
The strata can be classified as follows:
\begin{itemize}
\item[{\it i})]
$I_1=0$, $I_2=0$;
($x=0$, $j_1=0$, $j_2=0$)
The metric is singular and there is 
no spontaneous symmetry breaking. Therefore, the little group is
$SU(2)_X \otimes SU(2)_1 \otimes SU(2)_2 \otimes U(1)_X \otimes U(1)_J$.
The multiplets transform as (2,0,0,1,0) and (1,2,2,0,1).
\item[{\it ii})]
$I_2 = 0$;
($j_1=0$, $j_2=0$)
The metric is singular, and the little group is
$U(1) \otimes SU(2)_1 \otimes SU(2)_2 \otimes U(1)_J$.
The multiplets transform as (0,2,2,1), (0,1,1,0)
and (1,1,1,0).
\item[{\it iii})]
$I_1 = 0$, $I_4=I_2^2$;
($x=0$, $j_2=0$)
The metric is singular, and the little group is
$SU(2)_X \otimes U(1) \otimes U(1) \otimes U(1)_X$.
The  multiplets transform as (2,0,0,1), (1,0,-2,0),
(1,-1,1,0), (1,-1,-1,0) and (1,0,0,0).
\item[{\it iv})]
$I_1 = 0$, $I_4 = \frac{1}{2} I_2^2$;
($x=0$, $j_1=\pm j_2$)
The little group is $SU(2)_X \otimes SU(2) \otimes U(1)_X$.
The multiplets transform as (2,1,1),
(1,3,0) and (1,1,0).
\item[{\it v})]
$I_1=0$;
($x=0$)
The little group is $SU(2)_X \otimes U(1) \otimes U(1)_X$.
Two multiplets transform as (1,0,0), while
the remaining multiplets transform as (2,0,1), (1,-1,0) and
(1,1,0).
\item[{\it vi})]
$I_4 = I_2^2$;
($j_2=0$)
The metric is singular, and the little group is
$U(1) \otimes U(1) \otimes U(1)$.
Two multiplets transform as  (0,0,0),
while the remaining multiplets transform as (0,0,-2)
(1,0,0), (0,-1,1) and (0,-1,-1).
\item[{\it vii})]
$I_4=\frac{1}{2} I_2^2$;
($j_1=\pm j_2$)
The little group is $U(1) \otimes SU(2)$.
Two multiplets transform as (0,1), 
while the remaining multiplets transforms as (0,3) and (1,1).
\item[{\it viii})]
Generic $I_1$, $I_2$, $I_4$;
(generic $x_1$, $j_1$, $j_2$)
The little group is $U(1) \otimes U(1)$.
Three multiplets transform as (0,0), 
while the remaining multiplets transform
as (1,0), (0,-1) and (0,1).
\end{itemize}

$H_G$ orbits can be labeled by $\{I_1,I_2,I_3,I_4\}$,
or $\{x_1,x_2,j_1,j_2\}$, or $\{a,b,c,d\}$.
For each stratum, we indicate the subgroup of $H_G$ which remains unbroken, 
and also the remaining subgroup of the gauge group in case the gauge symmetry 
is not completely broken.
\begin{itemize}
\item[{\it i})]
$I_1=0$, $I_2=0$;
($x_1=0$, $x_2=0$, $j_1=0$, $j_2=0$)
The gauge and global symmetries remain unbroken. 
\item[{\it ii})]
$I_2=0$;
($x_2=0$, $j_1=0$, $j_2=0$)
The unbroken global symmetry group is 
$U(1) \otimes SU(2)_{\bar{F}} \otimes U(1)$.
The gauge symmetry is broken to $SU(3)$. 
The solution $T_1^{12}=a$, $\bar{F}_1^1=a$, $\bar{F}_2^2=a$, with
$x_1=2 a^3$, contains representative points of the
orbits in this stratum.
\item[{\it iii})]
$I_1=0$, $I_4=I_2^2$;
($x_1=0$, $x_2=0$, $j_2=0$)
The remaining global symmetry group is $U(1)\otimes U(1) \otimes U(1)$.
The gauge symmetry is broken to $SU(2)$.
The solution $T_1^{12}=a$, $ T_2^{45}=a$, $\bar{F}_4^1=a$, with
$j_1=12 a^4$ contains representative points of orbits
in this stratum.
\item[{\it iv})]
$I_1=0$, $I_4=1/2 I_2^2$;
($x_1=0$, $x_2=0$, $j_1=\pm j_2$)
The unbroken global symmetry group is $SU(2)\otimes U(1)$, and
the gauge symmetry is completely broken.
The solution $T_1^{12}=T_1^{34}=T_2^{15}=T_2^{24}=\bar{F}_1^1=\bar{F}_4^2=a$,
with $j_1=-j_2=12a^4$,
contains representative points of orbits in this
stratum.
\item[{\it v})]
$I_1=0$;
($x_1=0$, $x_2=0$)
The remaining global symmetry group is $U(1)\otimes U(1)$, and
the gauge symmetry is completely broken. 
The solution $T_1^{12}=a$, $T_1^{34}=T_2^{15}=\sqrt{a^2+b^2}$, $T_2^{24}=b$,
$\bar{F}_1^1=a$ and $\bar{F}_4^2=b$, with
$j_1=12a^2(a^2+b^2)$ and $j_2=-12b^2(a^2+b^2)$,
contains representative points of orbits in this
stratum.
The solution presented in Ref. \cite{ADS2}  also
contains representative points of orbits in this stratum.
\item[{\it vi})]
$I_3=I_1 I_2$, $I_4=I_2^2$;
($x_2=0$, $j_2=0$)
The unbroken global symmetry group is $U(1) \otimes U(1)$,
and the gauge symmetry broken to $SU(2)$.
The solution $T_1^{12}=a$, $T_1^{45}=T_2^{13}=b$, $\bar{F}_1^1=a$ and
$\bar{F}_2^2=\sqrt{a^2-b^2}$,
with
$x=2 a^2 \sqrt{a^2-b^2}$, $j_1=12a^2 b^2$ and $j_2=0$, 
contains representative points of orbits in this
stratum.
\item[{\it vii})]
$(2 I_3 - I_1 I_2)^2=I_1^2 (2 I_4 - I_2^2)$;
($x_2=0$)
The remaining global symmetry group is
$U(1)$, and the
gauge symmetry is completely broken. 
The flat directions presented in Refs. \cite{Shi,VZS}
contain representative points on the orbits in this stratum.
As shown in Ref. \cite{TtV2}, the classical vacuum of the
$SU(5)$ model with calculable supersymmetry breaking lies on
an orbit in this stratum with the property
$j_1=\pm j_2$. In terms of $H_G$ orbits, this additional condition
does not lead to a larger little group.
\item[{\it viii})]
$I_4^2=I_2^2$;
($j_2=0$)
The remaining global symmetry group is $U(1)$, and
the gauge symmetry is broken to $SU(2)$.
\item[{\it ix})]
Generic $I_1$, $I_2$, $I_3$, $I_4$;
(generic $x_1$, $x_2$, $j_1$, $j_2$)
Both global and gauge symmetries are completely broken. 
\end{itemize}
Even though every $H_G$ orbit is contained in an $H_M$ orbit, not
every stratum of $H_G$ orbits is completely contained in a stratum
of $H_M$ orbits. 

As in the $SU(3)$ model discussed in Section 1, 
non-perturbative effects completely change the classical
picture of the moduli space.
A non-perturbative effective
superpotential 
\beqa
W_{np} & = & \frac{\Lambda^{11}}{J_a^{\alpha} J_b^{\beta} 
                                 \epsilon_{\alpha \beta} \epsilon^{ab}},
\eeqa
generated by instantons, lifts the vacuum degeneracy
completely. 
However, instead of a mass term, which is not consistent
with the chiral nature of the $SU(5)$ theory, a renormalizable 
Yukawa--type interaction
in the superpotential can be introduced to stabilize the scalar potential. 
As described in Refs. \cite{ADS2,MV,TtV1,TtV2}, if the coupling constant
of this Yukawa term is sufficiently small, the theory below
the dynamical scale of the gauge interactions is a supersymmetric
sigma model, which has $X_a$ and $J_a^{\alpha}$ as coordinates,
$K_M$ as K\"{a}hler potential, and
\beqa
W & = & \frac{\Lambda^{11}}{J_a^{\alpha} J_b^{\beta} \epsilon_{\alpha \beta}
                       \epsilon^{ab}} + \lambda X_1
\eeqa
as  the superpotential.
In contrast to the $SU(3)$ model, the vacuum energy does not
vanish, and therefore supersymmetry is broken. The light 
mass spectrum, as calculated in Refs. \cite{TtV1,TtV2} displays
some degeneracies which can not be explained by the symmetry breaking
pattern of the global symmetry group of the gauge theory
including the superpotential.  However, as a consequence of the 
$H_M$ invariance of the K\"{a}hler potential, the symmetry group
of the sigma model extends the global symmetry group of the
full gauge theory.
In particular, the sigma model is invariant under
$SU(2)_{1} \otimes SU(2)_{2}$ transformations.
The degeneracies in the light spectrum square with
the breaking pattern of the extended symmetry
group of the sigma model.

\section{Conclusions}

We have presented a detailed study of the classical moduli space of the $SU(5)$ 
gauge theory with two anti--symmetric tensors and two anti-fundamentals.
We found that the symmetry group $H_M$ of the classical moduli space
extends the global symmetry group $H_G$ of the gauge theory. We analyzed
the moduli space in terms of orbits of both symmetry groups.

The extended symmetry of the moduli space has
two main consequences. Physically inequivalent classical vacua
of the gauge theory may have identical low energy limits, and
the effective models that describe the massless degrees of
freedom in the low energy limit have a symmetry group
that is larger than the unbroken subgroup of $H_G$. Even though
non--perturbative effects completely lift the classical 
moduli space, a remnant of the extended symmetry group 
of the K\"{a}hler potential
is the origin of degeneracies in the mass spectrum of the 
calculable $SU(5)$ model with dynamical supersymmetry breaking.

The extended symmetry of the classical moduli space is
traced to the fact that
the K\"{a}hler potential  does
not depend on an Hermitian invariant consistent
with the global symmetry group of the gauge theory.
We calculated the mass spectrum of the gauge theory for
vacua that are related by $H_M$ transformations but not
by $H_G$ transformations, and we found that
the mass spectrum of the massive vector multiplets differs.
This assured us that the additional symmetry of the moduli space
is not realized as a
symmetry of the full gauge theory. In fact, the same
evidence also eliminates the possibility that just the
scalar potential is invariant under the extended symmetry.

As an aside, the degeneracies in the spectrum of the massive
vector multiplets pose an intriguing question. In a
generic point of the moduli space, all global and gauge symmetries
are broken, and therefore no degeneracies are expected. However,
the existence of a degenerate quintuplet hints at some
kind of symmetry. 

Returning to the question of the extended
symmetry of the classical moduli space, we cannot completely
rule out the possibility that the full gauge theory, or the
just the scalar potential, is invariant under some symmetry
other than any of the extended symmetry transformations of
the classical moduli space, maybe even a discrete symmetry,
that we are unaware of. If such a symmetry exists and if it
forbids the absent terms in the K\"{a}hler potential, then the 
extended symmetry
of the classical moduli space that we have found would
be coincidental.

If the latter scenario is not realized, it is possible to take the 
point of view that the classical moduli spaces of 
supersymmetric chiral gauge theories with matter in tensor representations
have complicated structure, and that calculating their
K\"{a}hler potential provides an apt tool to understand this structure.
However, we find such a perspective somewhat unsatisfying and still
feel that it is worthwhile to seek
a fundamental principle that allows the determination
of the symmetries of the classical moduli space without
an explicit calculation of the K\"{a}hler potential.

Finally, we want to address the question whether the classical
moduli spaces of other supersymmetric gauge theories have
extended symmetries. 
Non-trivial flavor structure and matter transforming
under non-fundamental representations of the gauge group seem 
to be  prerequisites. However, with such matter content, the
parametrization of generic flat directions often is prohibitively complicated,
and an explicit calculation of the K\"{a}hler potential of the classical
moduli space is impossible.
This is particularly the case when the matter content is chosen so that
the gauge symmetry is non-anomalous, although this does
not seem to be required in a study of classical moduli spaces.

Looking at closely related models, the SU(5) model with one generation 
-- one anti-symmetric
tensor and one anti-fundamental -- has no flat directions. The
model with three generations has twenty-one moduli fields and its inequivalent
classical vacua are labeled by twenty-four parameters. Parametrizing
generic flat directions for this model is a forbidding task. Nevertheless,
the structure of the classical moduli space is of interest:
When non-perturbative effects are taken into account, 
the model is in an s-confining phase \cite{CSS}, and the structure of its 
classical moduli space is conjectured to be unmodified.

\section*{Acknowledgements} 
We thank Xerxes Tata for useful discussions, and Jeffrey Mandula
for stimulating comments concerning a possible connection
of our work with earlier studies of relativistic hydrogen atoms. 
The hospitality of the CERN Theory Division, where part of this research
was done, is gratefully acknowledged. 
This work was supported in part by the U.S. Department
of Energy under Grant No. DE-FG03-94ER40833.

\end{document}